# IMPROVING ON-TIME UNDERGRADUATE GRADUATION RATE FOR UNDERGRADUATE STUDENTS USING PREDICTIVE ANALYTICS

Ramineh Lopez-Yazdani [1], Roberto Rivera[1]

[1] Department of Mathematical Sciences, University of Puerto Rico-Mayaguez

## Abstract

The on-time graduation rate among universities in Puerto Rico is significantly lower than in the mainland United States. This problem is noteworthy because it leads to substantial negative consequences for the student, both socially and economically, the educational institution and the local economy. This project aims to develop a predictive model that accurately detects students early in their academic pursuit at risk of not graduating on time. Various predictive models are developed to do this, and the best model, the one with the highest performance, is selected. Using a dataset containing information from 24432 undergraduate students at the University of Puerto Rico at Mayaguez, the predictive performance of the models is evaluated in two scenarios: Group I includes both the first year of college and pre-college factors, and Group II only considers pre-college factors. Overall, for both scenarios, the boosting model, trained on the oversampled dataset, is the most successful at predicting who will not graduate on time.

# Introduction

The level of educational attainment of a population is vital for a country's productivity and economic prosperity. Since 2012, college enrollment in Puerto Rico has dropped from 245,495 to 184,921 (Rivera Molina, 2021). Factors such as the low high school graduation rate, about 51%, as well as steady decrease in younger population may be responsible for lower college enrollment. In addition to decreased enrollment, higher education institutes are faced with other challenges, such as the inability of most undergraduate students to graduate on time or even graduate at all. For the student, the inability to graduate has negative social, emotional, and financial consequences, including loss of potential income, accumulation of student loan debt and high default rate, and loss of professional opportunities with its direct negative effect on the local economy. Financial gains include lower student loan debt accruement and faster wealth growth because of earlier entry into the workforce. In addition, there is an 8% to 15% decrease in income in the ten years following graduation for those who did not graduate on time (Witteveen & Attewell, 2021). A 2016 report from the US Department of Labor reveals that for every extra year a student spends in college, the resulting median loss of income is $43,000 (Zumbrun, 2016). In 2014, it was reported that students could accrue $22,826 in debt for books, fees, transportation, and other college-related expenses for each extra year in college. Therefore, the total yearly financial loss is $68,153 (Rosenbaum J., 2014). This is compounded by the rapidly increasing cost of higher education (Abel & Deitz, 2014). College tuition increases faster than inflation so that every additional year will be more expensive than in previous years. Academic gain includes timely access to advanced studies, especially crucial for students seeking post-graduate studies, and a lower probability of college dropout. For example, according to the National Student Clearinghouse Research center, one-fourth of students drop out after four years (Kolodner, 2017). For the educational institution, delayed graduation translates into the loss of federal/state funding:

delays in college graduation translate into the university's inability to meet its annual quotas in accepting new students due to a lack of seats. Moreover, a low graduation rate may be perceived as a lack of faculty support and transparency, which could deter well-prepared and highly motivated students from attending that university (Steenhuyes & Vichery, 2022).

In this study, OTG is defined as the completion of the academic program at 100% of the time allotted in the curriculum. At the University of Puerto Rico-Mayaguez (UPRM) this means four years for non-engineering, and five years for engineering majors. Unfortunately, in Puerto Rico the OTG rate is significantly lower than for universities in the mainland US. For example, in 2020, the average OTG rate for the island was 8.8% (IPEDS, 2020). In other Latin and Caribbean countries, it has been reported that only 15% of undergraduate students graduate on time (Choque Dextre, 2015). In comparison, the mean OTG rate is higher in the US, at about 36% (IPEDS, 2011).

It is essential to correctly determine which students are at risk of not graduating on time and how to best help those students as early as possible. UPRM officials have already implemented a few mechanisms to help offset both the low OTG and overall graduation rate. These mechanisms include early access to stress management, study habits workshops, time management advice, and student help desk/tutoring. For example, help centers such as English Writing Center and CAT (*Centro Asistencia Tecnologia*) have enhanced students' professional writing skills and computer proficiency. Other resources currently provide students with social and academic tools to develop life and professional skills. Yet these mechanisms are available to students who take the initiative to seek the service; it would be helpful to identify which students require intervention in order to encourage them to access these and personalized services.

This research uses predictive analytics to examine available economic, social, academic, pre- and post-college data in order to predict whether undergraduate students at UPRM will graduate on time. The hope is that implementation of appropriate interventions can help increase OTG rates after detecting the students at risk of not graduating in time.

Predictive analytics is the process of analyzing available representative data to develop models which predict outcomes (not-OTG in our case, that is, those who will not graduate on time) with the highest degree of accuracy. It is an application of machine learning and is a widely used technique in business (Rivera, Principles of Managerial Statistics and Data Science, 2020), science (Manzorro, et al., 2022), epidemiology (Rivera, Rosenbaum, & Quispe, 2020; Rivera & Rolke, 2019; Lugo & Rivera, 2023; Rosenbaum, Stillo, Graves, & R., 2021) , and tourism (Rivera, 2016) to help individuals and organizations make educated decisions based on the predicted outcome.

A cohort of 24432 undergraduate students' records dating from 1999-2010 for the UPRM forms the dataset in this work. Student records have been provided by "*Oficina de Planificación Investigación y Manejamiento Institucional*" (OPIMI). Predictors include sociodemographic variables, high school academic achievement metrics, first-year college grade point average (GPA), school type, major, and faculty at the time of college admission.

In this work several supervised machine learning (ML) models are developed to predict who will not graduate on time. The models are supervised learners because they are trained on a dataset with known outcomes (Provost & Fawcett, 2013). The models which will be developed are: recursive Decision Tree, Random Forests, Gradient Boosting, Naive Bayes, Logistic Regression, and Attentive and Interpretable Tabular Learning (TabNet). Prediction assessment will be based on several evaluation metrics and the best model with the lowest classification prediction error is selected. The best model, with the lowest errors of prediction, will be selected.

In recent years, universities throughout the United States have utilized predictive analytic tools to boost their graduation rates. For example, since 2013, the University of Texas-Austin overall graduation rates have improved from 53% to 70% (Kash, 2019). This improvement is partially due to the early detection of behaviors not conducive to graduation and enabling these behaviors to be addressed by providing and encouraging access to guidance counselors and peer mentoring early in their college experience (Rolke, 2014). In addition, predictive analytics techniques have been used to improve the practical criteria to manage college enrollment and develop adaptive learning courseware. This has resulted in increased graduation rates (Ekowo & Palmer, 2016). Other universities have also deployed predictive analytic tools to improve graduation rates (Kash, 2019; Choque Dextre, 2015; Ekowo & Palmer, 2016; Hutt, Gardener, Kamentz, Duckworth, & D'Mello, 2018; Aiken, De Bin, & Hjorth-Jens, 2020; Barshay & Aslanian, 2019)

Among Hispanic students, other factors such as first-year student expectations and family support are also crucial in achieving a higher GPA which leads to a higher likelihood of on-time graduation. In other words, academic self-awareness with a positive outlook and high family graduation expectations are crucial (Creighton, 2007).

# Methods

## Description of Dataset

24432 undergraduate students' records dating from 1999-2010 for the University of Puerto Rico at Mayaguez (UPRM) forms the raw dataset in this project. 19545 student records (80% of original data) were used to train the machine learning models, and the remaining 4886 records (20% of original data) were utilized as the test subset. Assignments to the training or test subset were done at random. Table 1 contains information about predictors of interest and the target variable (OTG).

*Table 1 This table contains the variables, their descriptions and types. The number of categories are in parenthesis.*

| Variable | Description | Type |
|---:|---|---|
| Major | College major at the time of admission (54) | Categorical |
| Faculty | Faculty at the time of admission (5) | Categorical |
| Gender | Female or male (2) | Categorical |
| Year | Year of admission 1999-2010 (12) | Categorical |
| School_Type | Private, public, or others (3) | Categorical |
| Highschool.GPA | High school grade point average | Numeric |
| FAMILY_INCOME | Family income at the time of admission (5) | Categorical |
| Apt_Verbal | Verbal aptitude score | Integer |
| Aprov_Math | Mathematical achievement score | Integer |
| Apt_Math | Mathematical aptitude score | Integer |
| Aprov_Spanish | Spanish achievement score | Integer |
| Aprov_English | English achievement score | Integer |
| EDUC_FATHER | Mother's level of education (6) | Categorical |
| EDUC_MOTHER | Father's level of education- (6) | Categorical |
| 1_YR_GPA | First-year college GPA | Numeric |
| Rel_Stud_GPA | It is an index that measures the preparedness of the student across all high school. | Numeric |
| Rel_School_GPA | It measures how well the school has prepared the Student. | Numeric |
| OTG | Target: Y, success and N, failure (2) | Categorical |

.

This dataset is the product of merging three OPIMI raw datasets by matching student identifications. Family income was divided into five categories from high to low: higher than $50,000, $30,000-$49,999, $20,000-$29,999, $12,500-$19,999, and less than $12,499. Mother's

and father's education were divided into six categories: No education, less than high school, high school degree, Associate or less, College degree, and Graduate degree.

Numeric variables such as verbal aptitude, mathematical aptitude, mathematical achievements, Spanish achievements, and English achievements are scaled from 200 to 800.

The high school GPA does not uniformly gauge students' preparedness for college academic challenges across high schools. This minimizes predictive power of high school GPA. For example, a student with a high school GPA of 2.5 from high school <u>A</u> is not necessarily less academically prepared than another student with a GPA of 3 coming from school <u>B</u>. There are many reasons: degree of school academic standards/difficulty and the practice of grade inflation in some affluent schools are the most prominent. We create Relative student GPA to address this problem. It is a numeric variable that measures students' preparedness compared to other students across all high schools. It is calculated by dividing the student's high school GPA by the mean GPA of all admitted students from the same school. The higher the relative student GPA value is, the more prepared the student is in comparison to another student with a lower relative student GPA. For example, a student with a high school GPA of 3.2 graduating from a school A with a mean GPA of 3.0 (3.2/3.0=1.07) has excelled academically, better than average, compared to a student with the GPA of 3.3 from high school B with mean GPA of 3.5 (3.3/3.5=0.94), below average.

We also introduce relative school GPA which is a numeric metric that measures how well a high school has prepared the students to enter college. It compares the performance of high schools in terms of a student's first year in college performance. It is calculated by dividing the mean first-year in UPRM GPA of students who went to the same high school by means of the high school GPA of students from the same school. For example, the mean first-year UPRM GPA for the students from school A is 2.5, and high school mean GPA was 3.8. Therefore, for school A,

Rel_School_GPA (2.5/3.8) is 0.66. Students from school B with a 2.50 mean first year UPRM GPA but mean high school GPA of 3.2 has a Rel_School_GPA of 0.78. This means that even though school B has a lower mean high school GPA, students performed better as reflected in higher Rel_School_GPA (Rolke, 2014). There are 54 majors from which students can choose during the process of admission. Even though there are 4 faculties in UPRM, our dataset defines five faculties: Agricultural Sciences, Arts and Sciences - Art, Arts and Sciences - Sciences, Engineering, and Business Administration. The major is more descriptive than faculty but including this variable for Random Forest implementation becomes computationally infeasible. School types are: private, public, and other (it includes homeschooling).

**Machine learning algorithms**

In this project, ML models are supervised. The algorithms we deploy are: recursive Decision Trees; two ensemble tree-based models (Random Forest and Boosting); two probability-based models (Naïve Bayes and logistic Regression), and one artificial neural network for tabular data (TabNet). The predictive performance of these models will be evaluated upon training under minority class oversampled. Most of these machine learning models are well known (Hastie et al., 2009), except TabNet which we briefly describe.

### TabNet

The Deep Neural Network (DNN) has become a popular machine learning architecture, and duly so because of its robust classification performance for natural language processing (NLP), images, and speech recognition (Arik & Pfister, 2021). These data types are non-tabular, meaning that they are not organized in row and column format. Conversely, for tabular data DNN models are deficient in constructing decisions hyperplanes and have a high classification error rate.

To take advantage of the properties of both DNN and decision trees, the Google AI group has developed a newly modified DNN structure, TabNet. Unlike DNN, TabNet performs well even without inductive biases (Arik & Pfister, 2021). Even though the original research on TabNet has yet to be published in a peer reviewed journal, the preliminary research looks promising. Therefore, we decided to explore the TabNet structure and its predictive performance in this project.

TabNet is an attentive and interpretable algorithm. It incorporates decision tree (DT) hyperplanes and DNN architecture. Unlike DT, which relies on a hard threshold in the tabular setting, TabNet takes advantage of the soft and sequential attention method and chooses which features should be attended to at each decision step for every instance. This task is possible due to implementation of learnable sparse masks, vector of 0 and 1 as an output of sparsemax activation function, on the input features. In other words, it is an instance-wise sparse feature selection method of the most important features for each sample. Finally, the features are interpreted locally and globally, highlighting their importance individually for each sample and their contribution to the trained data (global) (Arik & Pfister, 2021).

There are two major components in TabNet, Encoder and Decoder, which we describe in what follows. The encoder architecture is a multi-step processor with N sequential decision steps. The $i^{th}$ step receives information from $(i-1)^{th}$ step and decides which features should be output and combined for the overall decision via a sparse learnable mask. Each step's vote is included in the final classification, comparable to an ensemble model. There are three main processes in each step: (i) Feature transformer - it processes the mask-selected features to generate the final output. It is a four-block and Gated Linear Unit (GLU) structure, where two are shared across every decision step, while the other two are independent. Each block has a fully connected layer (FC), a Batch Normalization (BN) layer, GLU; it is an activation function and introduces nonlinearity by multiplying sigmoid of input by input. The input is doubled, and one half is multiplied by a normalization factor $\sqrt{0.5}$, which helps stabilize the variance throughout the network, and then added to the other half. The output of the encoder unit splits into two equal subsets features. Rectified Linear Unit (ReLU) activation function is applied and the decisions of all the steps are summed. (ii) Attentive transformer - this unit is responsible for the learning of the sparse mask. It contains a block of three layers, FC, BN, and Sparsemax normalization (SN). Prior scales encode the use of each feature in the previous steps. There is a negative correlation between the number of features used in the previous step and the size of the prior weight. And finally, a sparsemax function that maps the result of multiplication of prior and feature transformer output (z) into a sparse learning sparse mask. (iii) Mask: points to the most important features by covering up features that are not considered important by the attentive transformer. It allows interpretability in the model both locally and globally.

The decoder architecture is a last step where the features are reconstructed by processing the encoded representation of feature transformer layer. The output is the map of the summation of all the steps selections after being passed through a FC layer.

Anaconda provides a platform for running the "pytorch_tabnet" application with a compatible scikit-learn interface and valuable methods for developing a TabNet learning model. Within this application exists the "pytorch_tabnet.tab_model" package, which contains a classifier class called TabNetClassifier (Arik & Pfister, 2021).

The data is divided into three subsets: training (80%), validation (10%), and testing (10%). The model is trained on the training subset, and the model performance is validated using the validation subset. In addition, the weights are optimized on the validation data set, and the training stops when the cross-entropy loss starts to increase on the validation set or when the early stop is invoked due to lack of cross-entropy loss improvement after a set number of epochs set by a hyperparameter called 'patience.' Torch.optim.Adam was utilized as a pytorch optimizer function for TabNet with a learning rate of 0.01.

### Classification Evaluation Metrics

Confusion Matrix (classification matrix) was used to evaluate the performance of each machine learning model. For the sake of this study, the true positive outcome has been defined as failure to graduate on time. Furthermore, we used: recall, F1-score, and the probability of incorrectly predicting a student graduated on time.

### Over-sampling

Our imbalanced dataset has a severely skewed class distribution- 11.2% graduated vs. 88.8% not graduated on time. The problem is that data imbalance can cause the machine learning classifier to become biased towards the majority class and ignore the minority class (graduation on time). As a result, the classifier will not be able to estimate a more accurate classification criterion. Specifically, the model may perform well in predicting not graduating on time but poorly in predicting students that do graduate on time. In this study, the positive class is the majority and over-sampling of minority class will provide us with a more balanced training dataset.

The process of over-sampling produces a new training dataset containing replicated observations from the minority class (1621 Y, and 12663 N instances). This is done by random resampling, with replacement, of the original minority observations in the training dataset. The new over-sampled training set contains 28568 instances with a more even target distribution, 14452 not graduated on time (Y=1) and 14116 (Y=0) graduated on time. In this project, we set the probability of minority class at 0.5. The test dataset was not oversampled and contained the original 3571 samples.

The R package ROSE (Randomly Oversampling Examples) was utilized to prepare an over-sampled training dataset (Lunardon, Menardi, & Torelli, 2014). The machine learning models were trained and fit to the over-sampled training set with 13227 N and 12925 data samples with the outcome of N and Y, respectively.

The ROSE method generates an artificial training dataset using kernel estimate of the conditional density $f(x|Y=y_i)$ where $y_i$ is 0 or 1, for both minority and majority classes (Yes or No). The kernel distribution function is centered on each data point, for both classes with diagonal covariance matrix of scaled parameters (Menardi & Torelli, 2010).

Missing values in the dataset can affect the performance of the predictive model. When discarding rows with at least one NA feature, relevant information goes unused. However, we found that missing value imputation (Van Vuuren & Groothuis-Outshoorn, 2011) did not significantly improve the predictive capabilities of our predictive models.

In this project, Jupyter Notebook platform was utilized to develop the TabNet model. It is a web-based open-source web application that utilizes the computer language Python (Python Software Foundation, 2001). The rest of the models were developed in RStudio IDE using R language (RStudio Team, 2020).

To handle demographic and academic data the completion certificate for "students conducting no more than minimal risk research" course was obtained from the office of Collaborative Institutional Training Initiative (CITI).

# Results

After the deletion of NAs, the dataset contained student records from 17855 students' samples and 18 features. First, the data was divided into 80% training and 20% test subsets: 14284 and 3571 students, respectively. Then, the exploratory analysis was performed using the training subset.

### Machine Learning Models

This project focuses on the performance of a set of machine learning models: recursive Decision Trees, Random Forest, Boosting, Naïve Bayes, logistic regression, and TabNet. The analysis proceeds as follows:

- Training of the models on a training dataset including first-year GPA and relative school GPA, the features related to the first-year performance (Group-I).
- Training of the models on the training dataset excluding first-year GPA and relative school GPA and considers only the pre-college features (Group-II).

For each scenario, the models are trained on two types of training datasets:

(1) all rows with missing values are deleted (original dataset),

(2) the training dataset without missing values (eliminated) over-sampled (over-s).

High Recall and F1-score values indicate the model's high capacity to predict students who will not graduate on time correctly. The preferred predictive model is the ML model with the highest Recall and F1-score values and the lowest misclassification error (probability of false-negative prediction).

*Table 2 Classification Evaluation Metrics - **Group – I**.*

| Metrics | RCDT | Random Forest | Boosting | Naïve Bayes | Logistic Reg. | TabNet |
|---|---|---|---|---|---|---|
| **Recall** | 0.9676 | 0.9852 | **0.9909** | 0.8154 | 0.9735 | 0.7370 |
| **F1-score** | 0.9367 | 0.9436 | 0.9449 | 0.8778 | **0.9432** | 0.833 |
| **P(Incorrectly Predicted as graduated)** | 0.0288 | 0.0131 | **0.0081** | 0.1643 | 0.0235 | 0.2341 |

In Group-I, all models trained on the original training dataset perform well, except for TabNet, with low Recall and F1-score values of 73% and 83%, respectively (Table 4.2). Boosting's performance is significantly better based on a Recall of 99.1% and an F1-score of 94.49%. This means that, out of students who did not graduate on time, it correctly classifies those who did not graduate on time 99.1% of the time. In addition, the misclassification error, incorrectly predicted as graduated, is the lowest for Boosting model, 0.8%. Naïve Bayes model has the second lowest Recall and F1-score values of 81% and 88%, respectively.

The predictive performance of all ML models was also evaluated after being trained on NA imputed training dataset. The predictive performance of the ML models was tested using a test dataset which contains 3590 students registers (see section 3.5). After the imputation of the missing values, the Recall performance of Boosting model improved slightly, by 0.19%, to 99.28%. Interestingly, the performance of the Logistic Regression model supersedes Boosting by 0.08%, based on the F1-score, to 94.75%. The misclassification error decreased to 0.6% for Boosting model, after imputation. The misclassification error for the TabNet model increased from 23.4% to 34.6% along with a lowering of Recall and F1-score values to 61% and 75%, respectively.

Without any exception, the classification performance of all predictive models worsened after training on an over-sampled train dataset. This is expected since oversampling is meant to provide more honest errors for an imbalanced dataset, not to reduce errors (Table 4.3). By oversampling, class classifications are more evenly distributed, minimizing the effects of potential biases for majority class prediction. In other words, predictive models do not automatically assign all new data samples to the majority class. It is noteworthy that Boosting model performance, both Recall, and F1-score, still exceeds the other models. For example, the Random Forest model with a Recall value of 84.4% lags the performance of Boosting model by almost 4%. RCDT model's predictive performance is the poorest, with 60% Recall and 74% F1-score and a higher misclassification error of 35%. Overall, Boosting model minimized the misclassification error- prediction error of on-time graduation when the student did not graduate- the most- to 15%. In comparison, imputation of missing values decreased this error to the lowest level of almost 0.6%.

*Table 3 Classification Evaluation Metrics - **Group – I**. The ML models are trained on both **original** and **over-sampled** (over-s) training datasets.*

| Metrics | RCDT | | Random Forest | | Boosting | | Naïve Bayes | | Logistic Reg. | | TabNet | |
|---|---|---|---|---|---|---|---|---|---|---|---|---|
| | Original | over-s | Original | over-s | Original | over-s | Original | over-s | Original | over-s | Original | over-s |
| **Recall** | 0.9676 | 0.6013 | 0.9852 | 0.7503 | **0.9909** | 0.8245 | 0.8154 | 0.6630 | 0.9735 | 0.7598 | 0.7370 | 0.7014 |
| **F1-score** | 0.9367 | 0.7427 | 0.9436 | 0.8439 | **0.9449** | 0.8812 | 0.8778 | 0.7884 | 0.9432 | 0.8510 | 0.833 | 0.8116 |
| **P(Incorrectly Predicted as graduated)** | 0.0288 | 0.3550 | 0.0131 | 0.2223 | **0.0081** | 0.1562 | 0.1643 | 0.3002 | 0.0235 | 0.2139 | 0.2341 | 0.2661 |

In Group-II, all models trained on the original train dataset also perform well, with TabNet showing the poorest performance, with a minimum Recall and F1-score values of 70% and 80%, respectively (Table 4). Boosting's performance is slightly better based on a Recall value of 99.7% and is tied with Logistic Regression for F1-score of 94.5%. In addition, the misclassification error is the lowest for Boosting model, 0.25%. After TabNet, Naïve Bayes models have the lowest sensitivity of 81% and 87%, respectively.

After the imputation of the missing values, the Recall value for Boosting model worsened slightly, by 1.43%, to 98.28%. Interestingly, the performance of RCDT model supersedes

Boosting by 1.47%, based on the Recall score, to 99.75%. In addition, the misclassification error was the lowest for imputed RCDT model, 0.2%. Overall, there is a noticeable lowering of Recall and F1-score in Boosting compared to those values from the Boosting model fitted on the original training dataset.

*Table 4 Classification Evaluation Metric - **Group– II**. The ML models are trained on both **original** and **imputed** training Datasets.*

| Metrics | RCDT | | Random Forest | | Boosting | | Naïve Bayes | | Logistic Reg. | | TabNet | |
|---|---|---|---|---|---|---|---|---|---|---|---|---|
| | Original | Imputed | Original | Imputed | Original | Imputed | Original | Imputed | Original | Imputed | Original | Imputed |
| **Recall** | 0.9515 | **0.9975** | 0.9930 | 0.9928 | 0.9971 | 0.9828 | 0.8160 | 0.8044 | 0.9874 | 0.9862 | 0.6976 | 0.7993 |
| **F1-score** | 0.9417 | 0.9448 | 0.9431 | 0.9434 | **0.9449** | 0.9442 | 0.8745 | 0.8688 | 0.9445 | 0.9459 | 0.8027 | 0.8868 |
| **P(Incorrectly Predicted as graduated)** | 0.0076 | **0.0022** | 0.0061 | 0.0064 | 0.0025 | 0.0064 | 0.1638 | 0.1746 | 0.0112 | 0.0122 | 0.2674 | 0.1705 |

Except for TabNet, all model classification performance metrics worsened upon training on an over-sampled train dataset (Table 5); this is similar to Group-I's results (Table 3). The Boosting model still performs the best; both Recall and F1-scores surpass the other models and false negative prediction, misclassification error, is lower. TabNet for model's Recall and F-score values improve from 70% to 72%, 80% to 83%, respectively. Misclassification error also improved to 25% from 27%. Moreover, the probabilistic models, Naïve Bayes and Logistic Regression perform similarly in all three metrics. Generally, these metrics are incredibly good for several models and may indicate a class imbalance, for which oversampling is correcting.

*Table 5 Classification Evaluation Metrics - **Group – II**. The models are trained on both **original** and **over- sampled** (over-s) training datasets.*

|  | RCDT | | Random Forest | | Boosting | | Naïve Bayes | | Logistic Reg. | | TabNet | |
|---|---|---|---|---|---|---|---|---|---|---|---|---|
| **Metrics** | Original | over-s | Original | over-s | Original | over-s | Original | over-s | Original | over-s | Original | over-s |
| **Recall** | 0.9515 | 0.5488 | 0.9930 | 0.7262 | **0.9971** | 0.8261 | 0.8160 | 0.6145 | 0.9874 | 0.7208 | 0.6976 | 0.7203 |
| **F1-score** | 0.9417 | 0.6974 | 0.9431 | 0.8219 | **0.9449** | 0.8829 | 0.8745 | 0.7504 | 0.9445 | 0.8039 | 0.8027 | 0.8271 |
| **P(Incorrectly Predicted as graduated)** | 0.0076 | 0.4018 | 0.0061 | 0.2439 | **0.0025** | 0.1548 | 0.1638 | 0.3344 | 0.0112 | 0.2486 | 0.2674 | 0.2493 |

While feature importance can be performed, such an analysis is often confused with a causative relationship with OTG. Feature importance only establishes the degree of relevance in predicting the output (under a given ML model) and thus is not presented here.

## Conclusion and Future Work

We developed robust and accurate predictive models used to improve the on-time graduation (OTG) rate for undergraduate students in UPRM. This task was accomplished by developing a predictive model with the greatest capacity to predict who will not graduate on time, using available features in the data set. To correct for class imbalance, we deployed an oversampling approach. Evidence was found of boosting models performing well in predicting on time graduation rates.

OTG rate analysis must be done as early as possible in the undergraduate's academic career to maximize the effectiveness of intervention programs. The model's predicted recommendations, followed by the immediate execution of new and already existing academic interventions, are hoped to improve the OTG rate in UPRM. However, the effectiveness of the model's implementation must be examined yearly.

It is important to emphasize that intervention recommendations are out of the scope of this study. According to the literature, interventions such as on-campus tutoring, career counseling, student-faculty, and peer mentoring services have prompted graduation rate improvement. For example, by providing at-risk students information about course requirements and on-campus resources, students are given a tool to evaluate their academic preparedness and decide whether they are mentally and academically ready to enroll in particular course (Sneyers & De Witte, 2018). In addition, enrolling in a course and dropping it later in the semester can extend the time of graduation, which negatively affects OTG (Nicholls & Gaede, 2014).

There are many aspects to a successful implementation of academic interventions. One crucial task is communicating effectively with the at-risk student via email, video call, or other preferred methods. In addition, frequent evaluation of the student participation and progress are valuable to the university officials, i.e., academic advisors and mentors, to adapt and customize methods to meet the student's needs. Finally, frequent written feedback from university officials allows students to monitor their progress. A pilot study will start soon at UPRM where undergraduate students determined by the model not to graduate on time are assigned to an intervention or a control group. Performance metrics such as annual student retention, mean GPA, and others will be contrasted among groups. Pending pilot study results the intervention program can be implemented campus wide and even across UPR campuses, though predictive models may need to be recalibrated across campuses.

One of the limitations of this study is that it does not consider the effect of students' transfer to another university or another program within UPRM on the OTG rate. In this project, a student is considered not to graduate on time even though they may have graduated on time somewhere else. However, it is unlikely that after a transfer, a student accomplished OTG. A government study shows that, on average, students who transfer lose 27 credits, in other words, one whole year of college credits (Stucker, 2022). Moreover, other factors such as student employment or work hours, commute time to school, and others may have a strong predictive power for OTG.

Future work should consider feature selection and engineering to improve the predictive models' performance and speed up the hyperparameter tuning process. This is done by eliminating the least important features, such as gender and school type, determined by their importance ranks.